\documentclass[nofootinbib]{revtex4}
\usepackage{graphicx}
\usepackage{fancyhdr}
\usepackage{amsmath,amssymb}
\usepackage{color}

\begin{document}

\title{Direct Probe of Majorana and Extended Higgs Particles \\
in Radiative Seesaw Models at the ILC}


\author{Hiroshi Yokoya}
\affiliation{Department of Physics, University of Toyama, Toyama
930-8555, JAPAN}

\begin{abstract}
A collider probe of the radiative seesaw models are considered.
Two key ingredients of these models, the extended Higgs sector and the
 source of the Majorana mass, although these details differ model by 
 model, would be studied at the TeV-scale electron-positron and
 electron-electron colliders.
The searches and mass determinations in the inert doublet model, which
 is the extended Higgs sector of the Ma model, are summarized as an
 example.
\end{abstract}

\maketitle

\thispagestyle{fancy}


%%%%%%%%%%%%%%%%%%%%%%%%%%%%%%%%%%
\section{Introduction}

The neutrino oscillation data clearly show that neutrinos have tiny
masses and large flavor mixing.
However, the standard model (SM) cannot explain them, while the masses
of the other SM bosons and ferminons are accounted for by the
Higgs mechanism. 
Therefore, after the discovery of a Higgs-boson at the LHC, the origin
of the neutrino masses and their tininess are imminent questions left to
us. 

The Majorana masses of the left-handed neutrinos are generated from 
the dimension-5 effective operators, 
\begin{align}
 {\cal L}=\frac{c_{ij}}{2\Lambda}\overline{\nu^c}^i_L\nu^j_L\phi^0\phi^0,
\end{align}
where $c_{ij}$ are dimensionless coefficients, $\Lambda$ is a cut-off
scale above which the internal structure of the vertex may be
resolved, and $\phi^0$ is the Higgs-boson field.
Through the vacuum expectation value of the Higgs boson $\langle
\phi^0\rangle$, the mass matrix for left-handed neutrinos is given as
$M^{ij}_{\nu}=c_{ij}\langle \phi^0\rangle^2/\Lambda$.
The tiny masses of neutrinos ($M^{ij}_{\nu}\lesssim0.1$~eV) mean
$c_{ij}/\Lambda\sim{\cal O}(10^{-14})$~GeV$^{-1}$ for $\langle
\phi^0\rangle\simeq256$~GeV. 
In the tree-level seesaw scenario, it is realized by introducing
right-handed neutrinos with Majorana masses of ${\cal O}(10^{14})$~GeV
and assuming $c_{ij}\sim{\cal O}(1)$. 
On the other hand, radiative seesaw models (RSMs) may be alternative
possibilities, in which the small coefficient $c_{ij}$ is realized via
the $n$-loop suppression factor
$\sim\left(1/16\pi^2\right)^n$~\cite{Zee:1980ai,Zee:1985id,Babu:1988ki,%
Krauss:2002px,Ma:2006km,Aoki:2008av,Ma:2008cu}.
As the results, the cut-off parameter $\Lambda$ can be much lower than
${\cal O}(10^{14})$~GeV, e.g.\ at the TeV scale.
Thus, in such models, direct searches and verification at collider
experiments may be possible. \\

General features of the RSMs are an extension of the Higgs sector and a
source of the Majorana mass, although their details are completely model
dependent. 
For example, in the Zee-Babu model~\cite{Zee:1985id,Babu:1988ki}, 
SU(2)$_L$ singlet scalars (which have non-zero lepton number) are
introduced with their lepton-number-violating interaction to generate
the Majorana masses of neutrinos at the two-loop level. 
The extended Higgs sector in the Ma model~\cite{Ma:2006km,Ma:2006fn} is
equivalent with the inert doublet model (IDM)~\cite{Barbieri:2006dq}
where one of the scalar doublet field is $Z_2$-odd, and that in the
Aoki-Kanemura-Seto~(AKS)
model~\cite{Aoki:2008av,Aoki:2009vf,Aoki:2011zg} is equivalent with the
two Higgs doublet model (THDM) with lepton-specific Yukawa
interactions~\cite{Aoki:2009ha}, with additional SU(2)$_L$ singlet
scalars which are odd under the $Z_2$-symmetry. 
As for the Majorana nature, TeV-scale right-handed neutrinos
are often introduced in various models, which are $Z_2$-odd to avoid the
tree-level Dirac mass. 
The introduced $Z_2$-symmetry stabilizes the lightest $Z_2$-odd particle
in the model.
Thus these models naturally contain a candidate of the dark matter. \\

Collider signatures of the RSMs also differ model by model.
To verify the model by experiments, we have to perform direct searches
of new particles at colliders, and if discovered, then by measuring
their properties and interactions we have to check whether these are
enough to describe the neutrino masses.
The collider searches of extended Higgs sector have been discussed
extensively~\cite{Gunion:1989we}. 
On the other hand, the Majorana nature can be probed by direct
searches of the TeV-scale Majorana
neutrinos~\cite{Han:2006ip,Kersten:2007vk}, or by observing the
lepton-number-violated processes, such like the neutrinoless double-beta
decay. 
The searches of right-handed neutrinos in the context of RSMs have been
studied in Refs.~\cite{Atwood:2007zza,Aoki:2010tf}.

%%%%%%%%%%%%%%%%%%%%%%%%%%%%%%%%%%
\section{Extended Higgs sector in the RSMs}

Collider signatures of the lepton-specific THDM are characterized by
multi-$\tau$
production~\cite{Aoki:2009ha,Kanemura:2011kx,Kanemura:2012az}, since the
extra scalars predominantly decay into $\tau$'s.
Such signatures can be observed at the LHC by requiring appropriate
cuts~\cite{Kanemura:2011kx}.
However, the mass determination would not be straightforward since the 
reconstruction of the $4\tau$ event kinematics cannot be solved.
On the other hand, the $4\tau$ kinematics can be easily
solved, and thus the masses and parameters in the model can be
determined well at the ILC~\cite{Kanemura:2012az}. 
The additional SU(2)-singlet scalars in the AKS model may be searched by
the energy scan of the production cross-sections of the charged-scalar
pair at the ILC~\cite{Aoki:2010tf}. \\

Collider signatures in the IDM are charged leptons plus missing
momentum, similar to those of charginos and neutralinos in
supersymmetric models. 
The searches of the inert doublet scalars at lepton and hadron colliders
have been studied in Refs.~\cite{Barbieri:2006dq,Cao:2007rm,%
Lundstrom:2008ai,Dolle:2009ft,Miao:2010rg,Gustafsson:2012aj}.
The signatures can be observed at the LHC only if the mass spectrum of
the scalars are in favorable cases. 
On the other hand, at the ILC, the inert doublet scalars can be easily
found, unless the masses are too heavy, and precise determinations 
of the masses and parameters are possible~\cite{Aoki:2013lhm}. 
In the next section, we summarize the detail studies of
the searches and mass determination of the additional scalars in the IDM
at the ILC based on Ref.~\cite{Aoki:2013lhm}.

%%%%%%%%%%%%%%%%%%%%%%%%%%%%%%%%%%
\section{Inert Doublet Model}

The IDM is one of the simplest extensions of the Higgs sector in the SM,
where an additional ${\rm SU}(2)_{L}$-doublet scalar field is
introduced, which is odd under the unbroken $Z_2$
symmetry~\cite{Barbieri:2006dq,Deshpande:1977rw}. 
Four kinds of additional scalars appear as physical states, namely
neutral $CP$-even state ($H$), neutral $CP$-odd state ($A$) and charged
scalar states ($H^{\pm}$), all of which are called inert scalars. 
Yukawa interactions of the inert scalars to SM fermions are forbidden
due to the $Z_2$ symmetry.
Because of the $Z_2$-parity conservation, the lightest inert particle
(LIP) becomes stable. 
Therefore, the model provides a scalar dark matter
candidate~\cite{Barbieri:2006dq,LopezHonorez:2006gr,Dolle:2009fn,%
Honorez:2010re,LopezHonorez:2010tb,Gustafsson:2012aj}.

The most general scalar potential can be written as 
\begin{align}
 V(\Phi_1,\Phi_2) &=
 \mu_1^2\left|\Phi_1\right|^2 + \mu_2^2\left|\Phi_2\right|^2
 + \frac{\lambda_1}{2}\left|\Phi_1\right|^4
 + \frac{\lambda_2}{2}\left|\Phi_2\right|^4
 + \lambda_3\left|\Phi_1\right|^2\left|\Phi_2\right|^2 \nonumber \\
 & + \lambda_4\left|\Phi_1^\dagger\Phi_2\right|^2
 + \left\{\frac{\lambda_5}{2}\left(\Phi_1^\dagger\Phi_2\right)^2
 + {\rm H.c.}\right\},
\end{align}
with seven real parameters $(\mu_1^2,\mu_2^2,\lambda_{1},
\lambda_{2},\lambda_{3},\lambda_{4},\lambda_5)$.
The potential has to satisfy theoretical constraints, such as the vacuum
stability~\cite{Deshpande:1977rw} and the 
perturbativity~\cite{Barbieri:2006dq}. 
By the vacuum stability at the tree level, the quartic terms are
constrained as $\lambda_1>0$, $\lambda_2>0$,
$\sqrt{\lambda_1\lambda_2}+\lambda_3>0$, and
$\sqrt{\lambda_1\lambda_2}+\lambda_3+\lambda_4-|\lambda_5|>0$~\cite{Deshpande:1977rw}. 
We consider the case where $\mu_1^2<0$,
$\lambda_1\mu_2^2>\lambda_3\mu_1^2$ and 
$\lambda_1\mu_2^2>(\lambda_3+\lambda_4+|\lambda_5|)\mu_1^2$ are
satisfied~\cite{Deshpande:1977rw}, so that $\Phi_2$ does not acquire the
vacuum expectation value (VEV) and only $\Phi_1$ plays a role of the
``Higgs-boson''. 
By denoting
\begin{align}
 \Phi_1= \left(\begin{array}{c}
	   0 \\ \frac{1}{\sqrt{2}}(v+h)
		\end{array}\right), \quad
 \Phi_2= \left(\begin{array}{c}
	   H^+ \\ \frac{1}{\sqrt{2}}(H+iA)
		\end{array}\right), 
\end{align}
where $v$ is the VEV, $v=\sqrt{-2\mu_1^2/\lambda_1}(\simeq246$~GeV), the
masses of these scalars are expressed as 
$m^2_h = \lambda_1 v^2$, $m^2_{H^+}=\mu_2^2+\frac{1}{2}\lambda_3 v^2$,
$m^2_{H}=\mu_2^2+\frac{1}{2}(\lambda_3+\lambda_4+\lambda_5)v^2$ and 
$m^2_{A}=\mu_2^2+\frac{1}{2}(\lambda_3+\lambda_4-\lambda_5)v^2$.
Thus, the seven parameters in the Higgs potential can be replaced by
the VEV $v$, four masses of the Higgs boson and inert scalars,
$(m_h,m_{H^+},m_{H},m_{A})$, the scalar self-coupling constant
$\lambda_2$, and $\lambda_H(\equiv\lambda_3+\lambda_4+\lambda_5)$ for
example.
To force the LIP to be electrically neutral, so that it can be a
candidate of the dark matter, $\lambda_4<|\lambda_5|$ must be
satisfied~\cite{Ginzburg:2010wa}. 
Depending on the sign of $\lambda_5$, either $H$ or $A$ becomes the
LIP.\footnote{%
In the Ma model, another possibility that the right-handed neutrino
becomes the dark matter is not excluded~\cite{Kubo:2006yx}.
In such cases, constraints on the parameters, $\lambda_4$ and
$\lambda_5$, are relaxed.}
Hereafter, we take $H$ as the LIP. 

\begin{table}[b]
 \begin{tabular}{c||ccc|cc}
  & \multicolumn{3}{c|}{Inert scalar masses} &
  \multicolumn{2}{c}{ILC cross sections [$\sqrt{s}=250$~GeV (500~GeV)]} \\
  & $m_{H}$~[GeV] & $m_{A}$~[GeV] & $m_{H^\pm}$~[GeV] &
  $\sigma_{e^+e^-\to HA}$~[fb] & $\sigma_{e^+e^-\to H^+H^-}$~[fb] \\
  \hline
  (I)   & 65. & 73.  & 120. & 152. (47.) & 11. (79.) \\
  (II)  & 65. & 120. & 120. & 74. (41.)  & 11. (79.) \\
  (III) & 65. & 73.  & 160. & 152. (47.) & 0. (53.) \\
  (IV)  & 65. & 160. & 160. & 17. (35.)  & 0. (53.)
 \end{tabular}
 \caption{Masses of inert scalars and ILC cross sections
 for our four benchmark points.}\label{tab:ilc}
\end{table}
For the collider study, four benchmark points for the masses of inert
scalars listed in Table~\ref{tab:ilc} are considered, which satisfy all
the available theoretical and also phenomenological
constraints~\cite{Gustafsson:2012aj}.
We study the case where the masses of $H$ and $A$ are close to each
other (I, III) for which the LEP and LHC experiments can not
probe~\cite{Lundstrom:2008ai,Cao:2007rm,Dolle:2009ft,Miao:2010rg,%
Gustafsson:2012aj}.
The other cases are when $m_A-m_H$ is medium (II) or large such that
the $Z$-boson from $A\to HZ$ becomes on-shell (IV).
For the $W$-bosons in $H^\pm\to W^\pm H$, we consider the
off-shell (I, II) and on-shell (III, IV) cases. 
For the four benchmark points, the production cross sections of inert
scalars at the ILC are large enough to be observed.
In Table~\ref{tab:ilc}, we list the cross sections of $HA$
production and $H^+H^-$ production at $\sqrt{s}=250$~GeV
and 500~GeV.

For the cases (II, IV), $H^{\pm}$ decays into $W^\pm H$ predominantly,
where we admit the $W$-boson to be off-shell if $m_{H^{\pm}}-m_{H}<m_W$. 
While for the cases (I) and (III), $H^{\pm}\to W^\pm A$ decay would be
sizable as well, with the branching ratios about 32\% and 27\%,
respectively.
The decay of the $A$-boson is dominated by $A\to Z^{(*)}H$. \\

Collider signatures of the inert scalars in the IDM have been studied in
the literature~\cite{Barbieri:2006dq,Cao:2007rm,Lundstrom:2008ai,%
Dolle:2009ft,Miao:2010rg,Gustafsson:2012aj}.
In Ref.~\cite{Lundstrom:2008ai}, bounds on the masses of the inert
scalars are obtained by using the the LEP II data.
Even though the parameter regions where the inert scalars could be
discovered at the
LHC are pointed out~\cite{Dolle:2009ft,Miao:2010rg,Gustafsson:2012aj},
detailed analysis on these scalars such as the precise 
determination of these masses and quantum numbers would be performed at
lepton colliders. 

\subsubsection*{$e^+e^-\to H^+H^-$ process}

Here the $H^+H^-$ pair production at the ILC, where $H^\pm$
predominantly decays into $HW^{\pm}$, and $W^{\pm}$ further into
$\ell^{\pm}\nu$ or $q\bar{q}'$ are studied. 
The semi-leptonic and all-hadronic decay modes are used as successful
signatures.

First, we study the semi-leptonic decay mode, where the signature is
$\ell^\pm jj$ plus large missing energy. 
The leading background process would be $\tau^\pm\nu jj$ production
followed by the leptonic decay of $\tau$. 
The $\ell^\pm\nu jj$ background process can be reduced by requiring
a large recoil mass.
The contribution from production of $\mu^+\mu^-jj$ and missing
particles, where one of the muons goes out of the acceptance region, are
negligible. 
The event simulation for the case (I) [the case (III)] is only differ
from that for the case (II) [the case (IV)] by the overall
normalization. 

\begin{figure}[tb]
 \begin{center}
  \includegraphics[width=0.3\textwidth,clip]{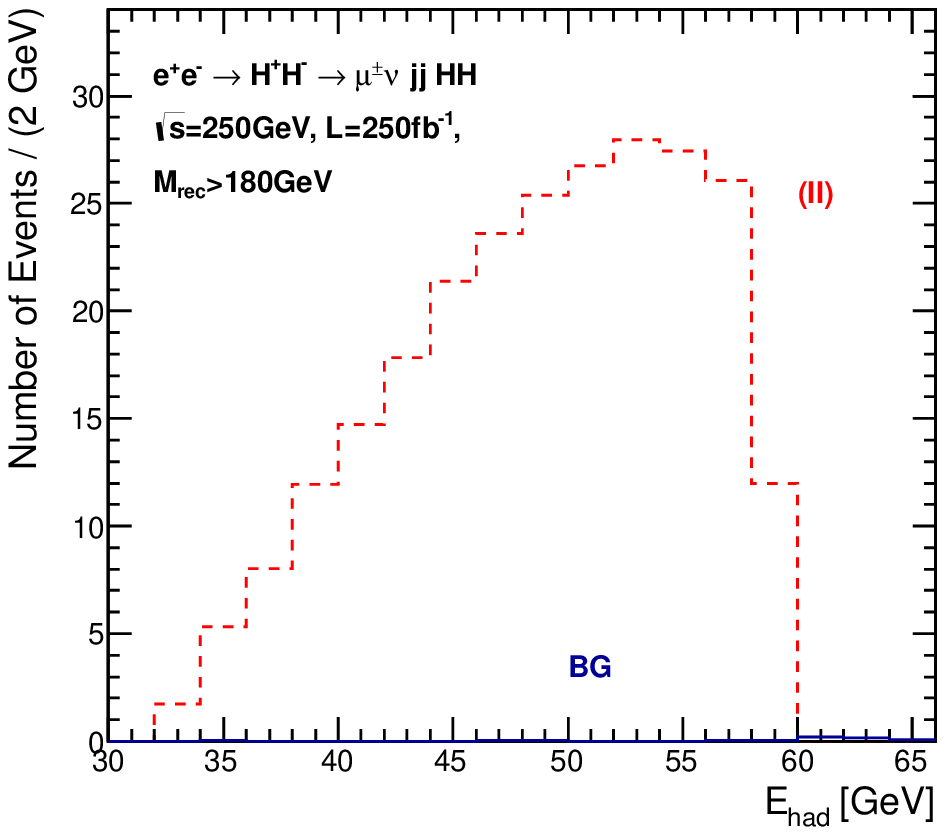}
  \includegraphics[width=0.3\textwidth,clip]{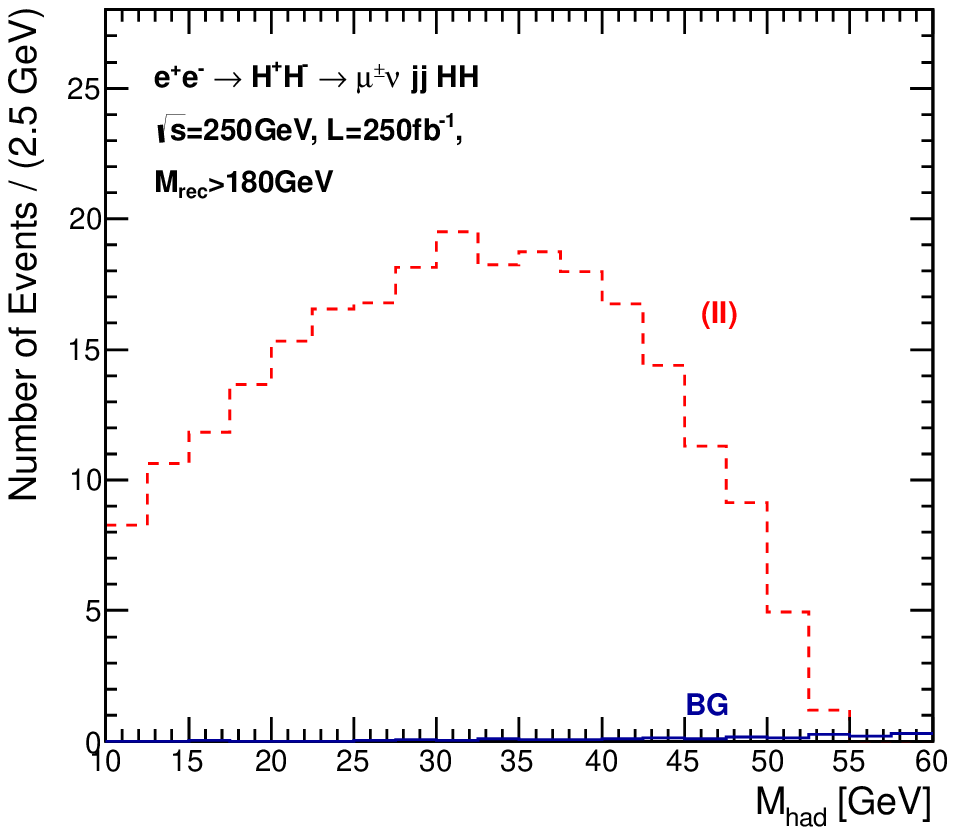}
  \includegraphics[width=0.3\textwidth,clip]{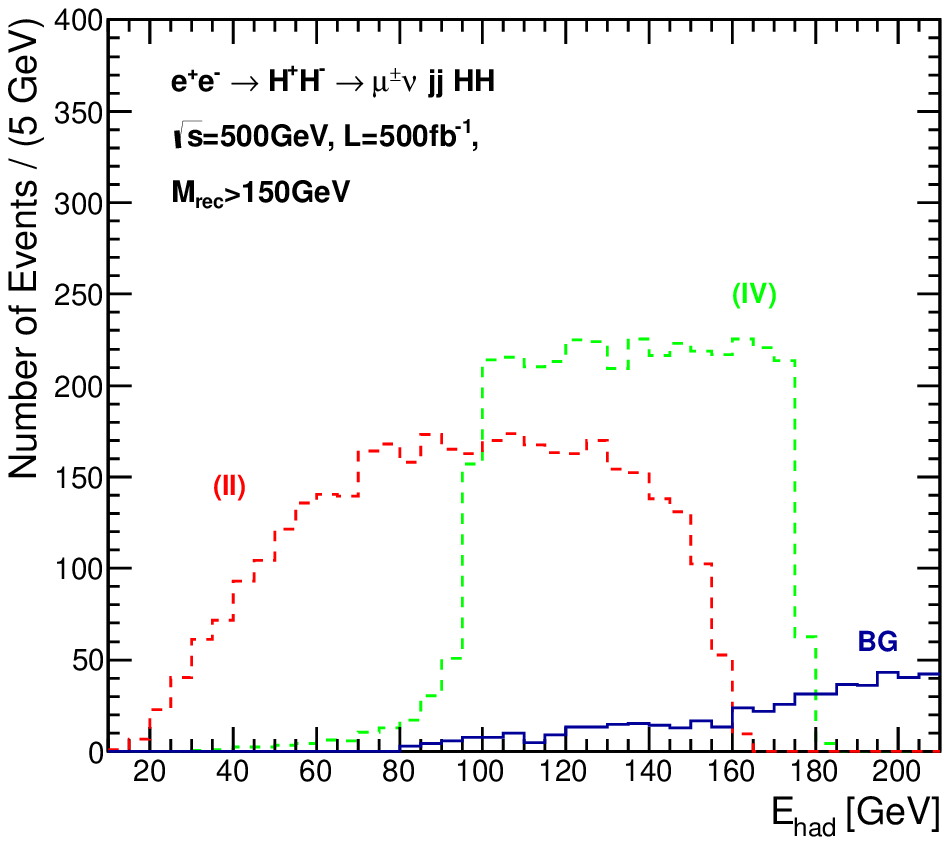}
  \caption{Distributions of $E_{\rm had}$, $M_{\rm had}$ in the
  semi-leptonic decay mode at $\sqrt{s}=250$~GeV with ${\cal L}_{\rm
  int}=250$~fb$^{-1}$ (left and middle) and
  that of $E_{\rm had}$ at $\sqrt{s}=500$~GeV with ${\cal L}_{\rm
  int}=500$~fb$^{-1}$ (right).}
  \label{fig:Semi}
 \end{center}
\end{figure}
In the left and middle panels in Fig.~\ref{fig:Semi}, $E_{\rm had}$ and
$M_{\rm had}$ distributions in the semi-leptonic decay mode are plotted
by using the parameter set (II) at the ILC with $\sqrt{s}=250$~GeV and
${\cal L}_{\rm int}=250$~fb$^{-1}$ with a cut of $M_{\rm rec}>180$~GeV. 
For the case with the off-shell $W$-boson, the endpoints of the all-jets
(hadrons) energy distribution are given by
\begin{align}
 E_{\rm had}^{\rm max/min} =
 \frac{\sqrt{s}}{4}\left(1-\frac{m_{H}^2}{m_{H^\pm}^2}\right)
\left[1\pm\sqrt{1-\frac{4m_{H^\pm}^2}{s}}\right]. \label{eq:1}
\end{align}
Furthermore, the maximum value of the invariant mass of all hadrons is
just the difference between $m_{H^\pm}$ and $m_{H}$, 
\begin{align}
M_{\rm had}^{\rm max}=m_{H^\pm}-m_{H}. \label{eq:2}
\end{align}

In the right panel in Fig.~\ref{fig:Semi}, the $E_{\rm had}$
distribution in the semi-leptonic decay modes are plotted by using the
parameter sets (II) and (IV) at the ILC with $\sqrt{s}=500$~GeV and
${\cal L}_{\rm int}=500$~fb$^{-1}$ with a cut of $M_{\rm rec}>150$~GeV. 
Notice that the parameter set (II) corresponds to the case where
$H^{\pm}$ decays into off-shell $W$ and $H$, and (IV) corresponds to the
case where $H^{\pm}$ decays into on-shell $W$ and $H$.
When the $W$-boson is on-shell, the signal distribution is like a
rectangle where the edges are given by
\begin{align}
 E^{\rm max/min}_{\rm had} = \gamma_{H^\pm}\hat{E}_{\rm had} 
 \pm \gamma_{H^\pm}\beta_{H^\pm}\hat{p}_{\rm had},\label{eq:4}
\end{align}
with $\gamma_{H^{\pm}} = \sqrt{s}/(2m_{H^\pm})$,
 $\beta_{H^{\pm}} = (1-4m_{H^\pm}^2/s)^{1/2}$, $\hat{E}_{\rm
 had} = (m_{H^\pm}^2-m_{H}^2+m_W^2)/(2m_{H^\pm})$ and
 $\hat{p}_{\rm had} =
 m_{H^\pm}/2\times\lambda(1,m_{H}^2/m_{H^\pm}^2,m_W^2/m_{H^\pm}^2)$.

Then, we present the all-hadronic decay mode, which gives the four jets
plus large missing energy signatures. 
Main SM background contributions are the production of four partons with
two neutrinos. 

\begin{figure}[t]
 \begin{center}
  \includegraphics[width=0.3\textwidth,clip]{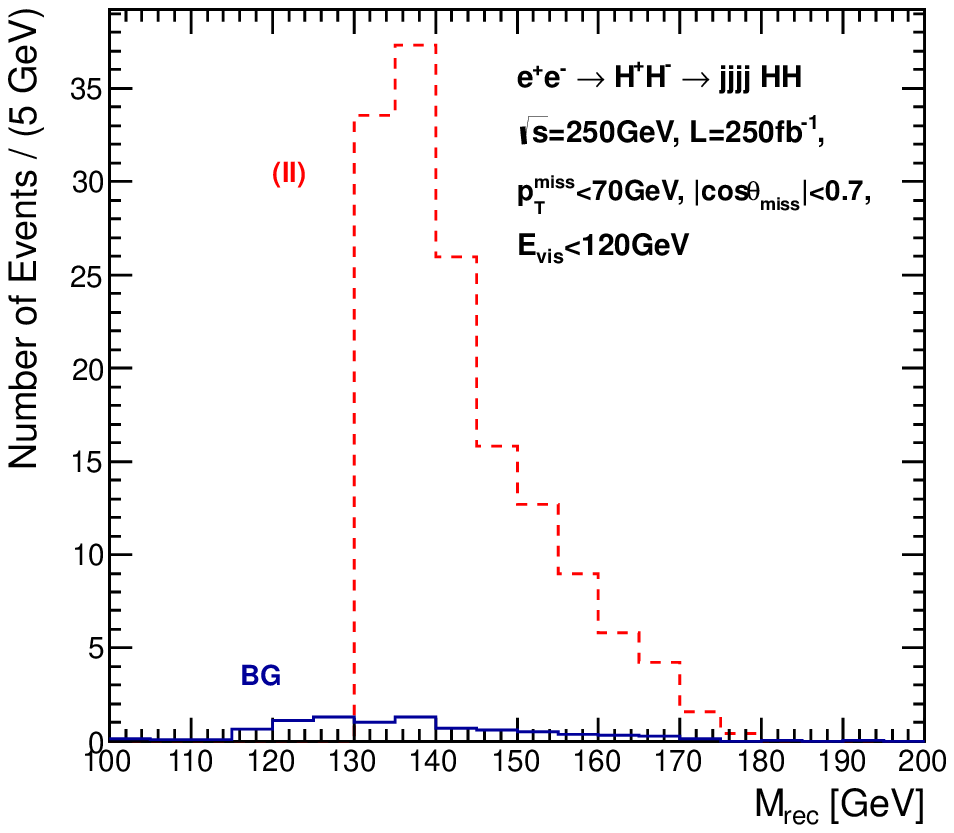}
  \includegraphics[width=0.3\textwidth,clip]{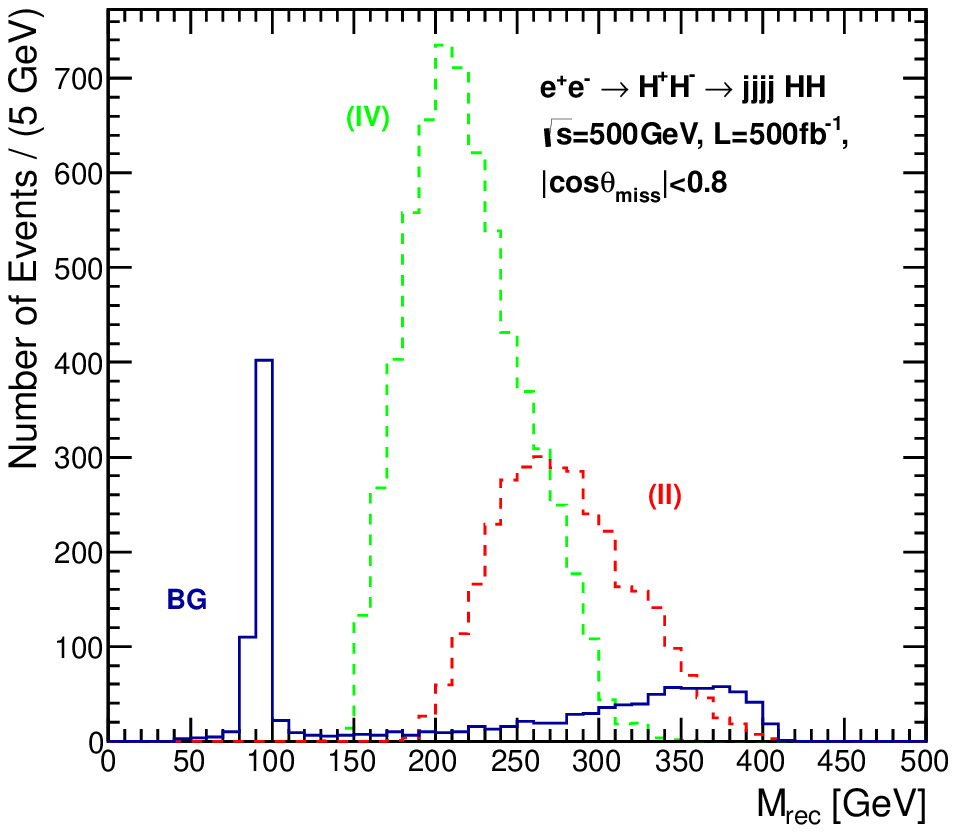}
  \includegraphics[width=0.3\textwidth,clip]{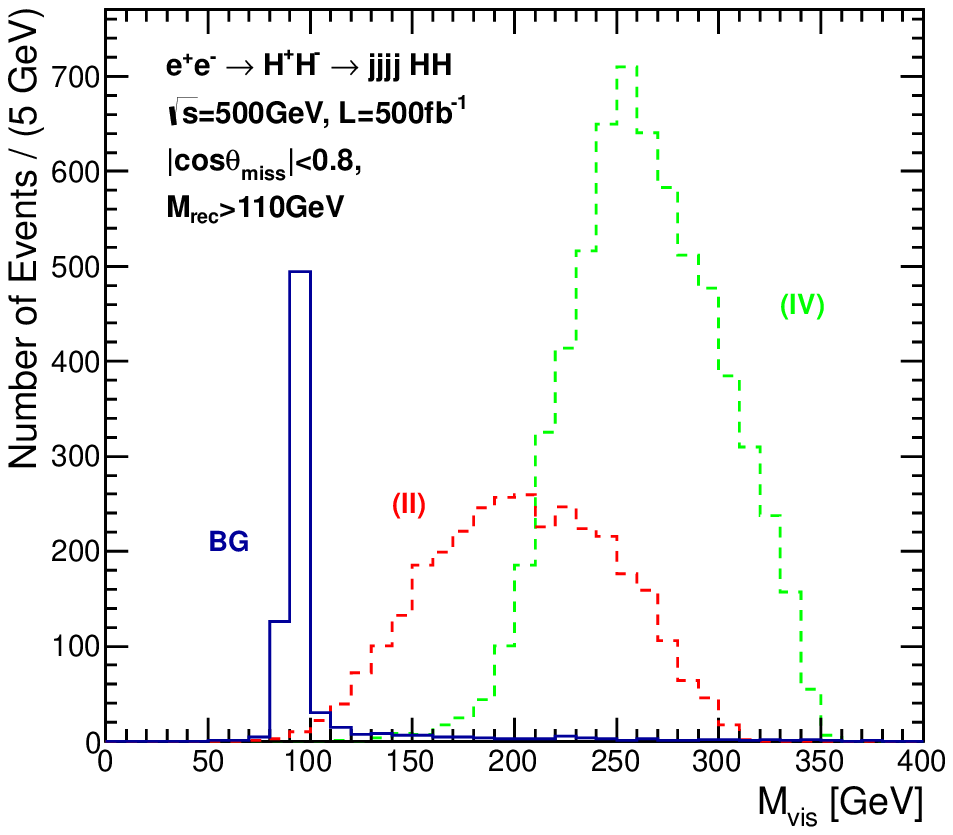}
  \caption{Distributions of $M_{\rm rec}$ in the all-hadronic decay
  mode at $\sqrt{s}=250$~GeV with ${\cal L_{\rm
  int}}=250$~fb$^{-1}$ (left), and $M_{\rm rec}$ and $M_{\rm 
  vis}$ distributions in the all-hadronic mode at $\sqrt{s}=500$~GeV
  with ${\cal L_{\rm int}}=500$~fb$^{-1}$ (middle and right).}
  \label{fig:Dist_On}
 \end{center}
\end{figure}

In the left panel of Fig.~\ref{fig:Dist_On}, $M_{\rm rec}$ distribution
is plotted for the signal process using the parameter set (II) at
$\sqrt{s}=250$~GeV with ${\cal L}_{\rm int}=250$~fb$^{-1}$. 
To reduce the SM background, kinematical cuts of $p_{T}^{\rm
miss}>70$~GeV, $|\cos\theta_{\rm miss}|<0.7$ and $E_{\rm vis}<120$~GeV
are applied, where $\theta_{\rm miss}$ is the polar angle of the missing
3-momenta and $E_{\rm vis}$ is the sum of the energy of all hadrons in
one event. 
As a result, the SM background is sufficiently reduced. 
The minimum of the $M_{\rm rec}$ distribution is at the twice of $m_H$, 
\begin{align}
 M_{\rm rec}^{\rm min}=2m_{H}.\label{eq:recmin}
\end{align}
In the middle panel of Fig.~\ref{fig:Dist_On}, the same distributions
are plotted but for the signal processes using parameter sets (II) and
(IV) at $\sqrt{s}=500$~GeV with ${\cal L_{\rm int}}=500$~fb$^{-1}$.
By the kinematical cut of $|\cos\theta_{\rm miss}|<0.8$, the SM
background is sufficiently reduced except at $M_{\rm rec}\simeq m_{Z}$.
The peak of the signal distribution is given by 
\begin{align}
 M_{\rm rec}^{\rm peak} = \frac{m_{H}\sqrt{s}}{m_{H^{\pm}}}.
\label{eq:recpeak}
\end{align}
This relation holds even when the $W$-boson in $H^\pm\to W^\pm H$ is
off-shell~\cite{Aoki:2013lhm}.
Thus, the ratio of $m_{H}$ and $m_{H^{\pm}}$ can be determined.

In the right panel of Fig.~\ref{fig:Dist_On}, the $M_{\rm vis}$
distributions are plotted for the signal processes using parameter
sets (II) and (IV) at $\sqrt{s}=500$~GeV with ${\cal L_{\rm
int}}=500$~fb$^{-1}$. 
In addition to the kinematical cut applied in the previous panel,
the cut of $M_{\rm rec}>110$~GeV is applied to reduce the SM background
with $Z\to\nu\bar\nu$. 
After these cuts, the SM background is sufficiently reduced except at
$M_{\rm vis}\simeq m_Z$. 
The signal distribution has a peak at 
\begin{align}
 M_{\rm vis}^{\rm peak}=\frac{m_{W}\sqrt{s}}{m_{H^{\pm}}},
\label{eq:vispeak}
\end{align}
when the $W$-boson in $H^\pm\to W^\pm H$ is on-shell [the case
(IV)]. 
When the $W$-boson is off-shell, the relation on the peak position no
more holds.

\begin{figure}[t]
 \begin{center}
  \includegraphics[width=0.4\textwidth,clip]{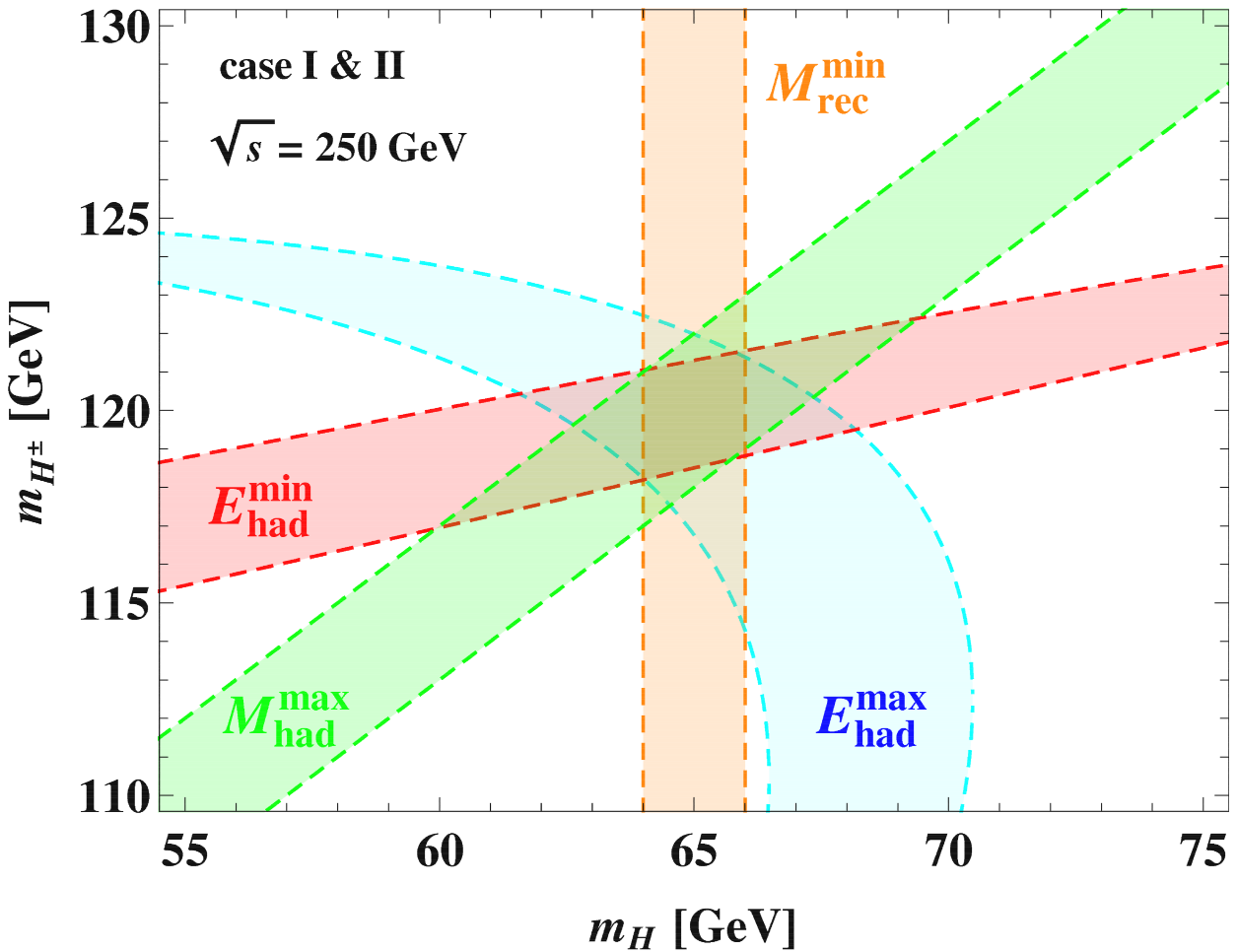}
  \hspace{30pt}
  \includegraphics[width=0.4\textwidth,clip]{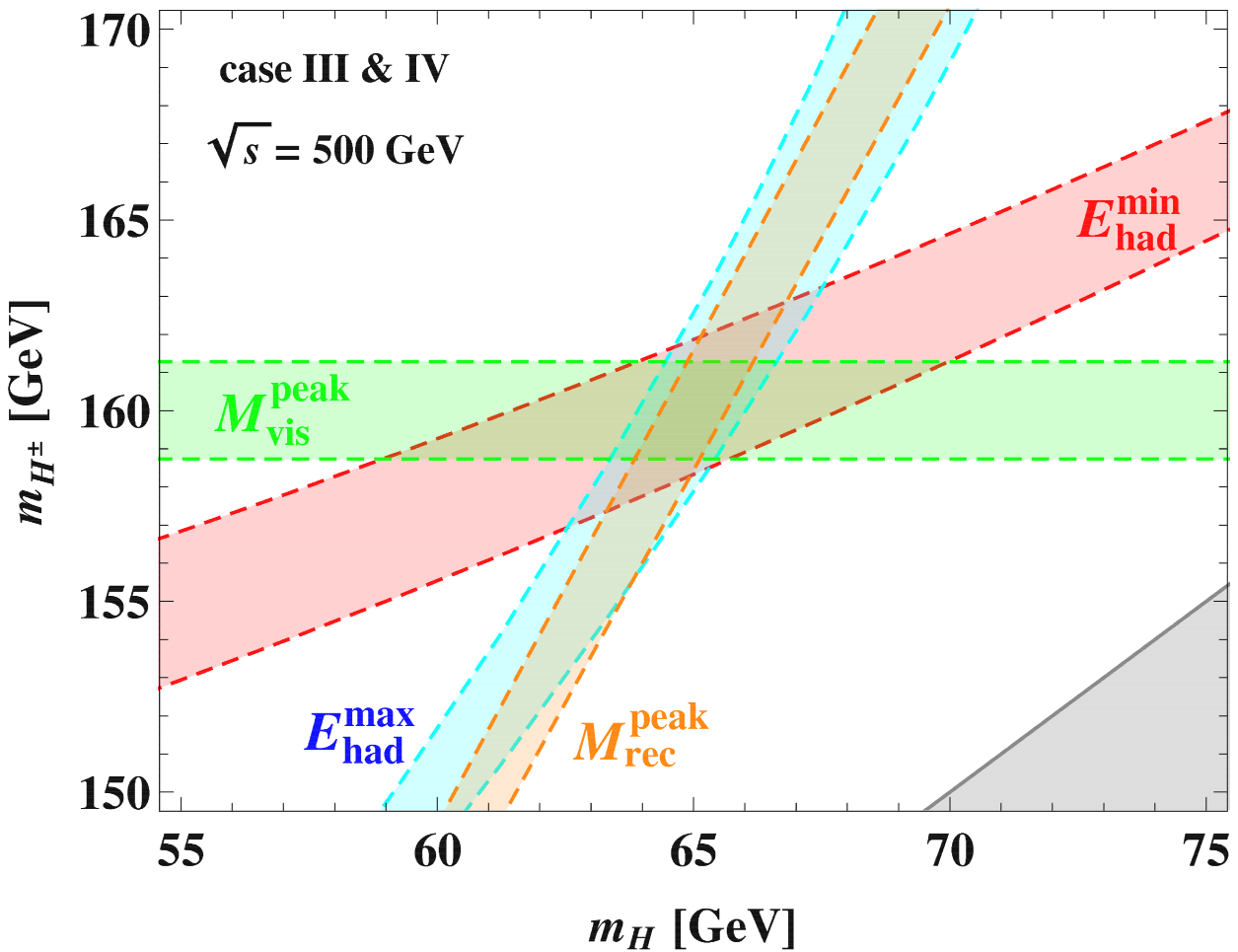}
  \caption{Determinations of $m_{H^\pm}$ and $m_{H}$ by the four
  observables are illustrated in the left [right] panel for the cases
  (I, II) [(III, IV)] at $\sqrt{s}=250$~GeV [500~GeV].
  Each observable is assumed to be measured in $\pm 2$~GeV accuracy.
  } \label{fig:Mass}
 \end{center}
\end{figure}

The observables for determining $m_{H^{\pm}}$ and $m_{H}$ in the process
$e^+e^-\to H^+H^-$ are summarized in Fig.~\ref{fig:Mass}. 
In the left panel, for $m_{H^{\pm}}-m_{H}\le m_{W}$, the the four bands
are plotted on the $m_{H^\pm}$-$m_{H}$ plane by assuming that the four
quantities, $E^{\rm max}_{\rm had}$, $E^{\rm min}_{\rm had}$ in
Eq.~(\ref{eq:1}), $M^{\rm max}_{\rm had}$ in Eq.~(\ref{eq:2}) and
$M^{\rm min}_{\rm rec}$ in Eq.~(\ref{eq:recmin}), are measured in $\pm
2$~GeV accuracy. 
For this assumption, the accuracy of the $m_{H^\pm}$ ($m_{H}$)
determination would be $\pm 2$~GeV ($\pm 1$~GeV).
On the other hand, if $m_{H^{\pm}}-m_{H}\ge m_{W}$, the four
observables, $E_{\rm had}^{\rm max}$, $E_{\rm had}^{\rm min}$ in
Eq.~(\ref{eq:4}), $M_{\rm rec}^{\rm peak}$ in Eq.~(\ref{eq:recpeak}) and
$M_{\rm vis}^{\rm peak}$ in Eq.~(\ref{eq:vispeak}) are utilized for the
mass determination. 
In the right panel of Fig.~\ref{fig:Mass}, the four bands are plotted on
the $m_{H^\pm}$-$m_{H}$ plane by assuming that the four
observables are measured in $\pm 2$~GeV accuracy. 
It turns out that the constraints from measurements of $M^{\rm
peak}_{\rm vis}$ and $M^{\rm peak}_{\rm rec}$ are more stringent than
those from the $E^{\rm max/min}_{\rm had}$ measurements, if these
quantities are measured in an equal accuracy. 
It is expected that peak positions can be precisely determined more than
endpoints of distributions in the presence of the resolution of energy
measurements and the remaining background contributions. 
By combining the four measurements with the uncertainty of $\pm 2$~GeV,
$m_{H^\pm}$ and $m_{H}$ can be determined in $\pm 1$~GeV accuracy. \\

The discovery of $A$-boson, the $CP$-odd inert scalar, is achieved by
using the $e^+e^-\to HA$ process followed by $A\to HZ$ decay.
The study of the collider signature and mass determination in this
process can be found in Ref.~\cite{Aoki:2013lhm}.

%%%%%%%%%%%%%%%%%%%%%%%%%%%%%%%%%%%%%%
\section{Majorana nature in the RSMs}

Sometimes, lepton colliders have bigger potential for the direct
searches of new particles in the RSMs than hadron colliders. 
Not only that, to determine the masses and parameters in the model,
precise measurements at lepton colliders would be more advanced, since
reconstruction of the recoil mass, decay angular distributions and
the energy scan of the production cross-sections are possible there.
In this report, we have seen that the extended Higgs sector of the
RSMs, e.g.\ IDM or the lepton-specific THDM, can be probed precisely at
the ILC. 

Furthermore, the Majorana nature in the RSMs can be probed as well at
lepton colliders~\cite{Atwood:2007zza,Aoki:2010tf}. 
The $t$-channel Majorana neutrino exchange diagrams have characteristic
contribution to the total cross-section and the angular distributions of
the produced particles and their decay products, even though the mass
of the $t$-channel particle is heavier than the scattering energy. 
An electron-electron collider option at the future ILC experiment would
deserve special attention for the direct probe of the
RSMs~\cite{Rizzo:1982kn,Heusch:1993qu,Belanger:1995nh}. 
Since the Feynman diagrams of the $e^-e^-$ scattering in the RSMs can be
regarded as parts of the diagrams for the neutrino mass generation, it
provides the direct test of the Majorana nature in the RSMs.

%%%%%%%%%%%%%%%%%
\section{Summary}

To summarize, the direct probe of the RSMs at colliders are studied.
The key ingredients of the RSMs are the extended Higgs sector and the
source of the Majorana mass, while the details are completely model
dependent.
The collider signatures also differ model by model.
It is observed that at lepton colliders not only the extended Higgs
sector but also the Majorana nature in the RSMs can be probed.

\bigskip % extra skip inserted
%%%%%%%%%%%%%%%%%%%%%%%%%%%%%%%%%%
\begin{acknowledgments}
The Author would like to thank Mayumi Aoki and Shinya Kanemura for
fruitful collaborations.
The work was supported in part by Grant-in-Aid for Scientific Research,
No.\ 24340036.
\end{acknowledgments}

\bigskip % extra skip inserted

\end{document}